\documentclass[numbook, envcountsect, envcountsame, envcountreset, runningheads, smallextended]{svjour3}

\usepackage{amscd,amsfonts,amssymb,amsmath,latexsym,array,hhline,xcolor,graphicx}

\usepackage{latexsym}
\usepackage{times}

\newcommand\cA{{\cal A}}

\newcommand\cL{{\cal L}}

\newcommand\cZ{{\cal Z}}

\def\E{{\bf E}}

\def\bbr{{\mathbb R}}

\def\text#1{\hbox{#1}}

\def\E{{\bf E}}

\def\build #1_#2{\mathrel{\mathop{\kern 0pt #1}\limits_{#2}}} 
\newcommand{\zs}[1]{{\mathchoice{#1}{#1}{\lower.25ex\hbox{$\scriptstyle#1$}}
{\lower0.25ex\hbox{$\scriptscriptstyle#1$}}}}

\numberwithin{equation}{section}

\def\bbr{{\mathbb R}}

\def\bbr{{\mathbb R}}
\newcommand\fdem{$\Box$}
\newcommand\beq{\begin{equation}}
\newcommand\eeq{\end{equation}}
\newcommand\bea{\begin{eqnarray}}
\newcommand\eea{\end{eqnarray}}
\newcommand\bean{\begin{eqnarray*}}
\newcommand\eean{\end{eqnarray*}}

\newcommand\beal{\begin{alighn}}
\newcommand\eeal{\end{align}}

\newtheorem{theo}{Theorem}[section]

\newtheorem{lemm}[theo]{Lemma}

\begin{document}

\title{ Optimal pair trading: consumption-investment problem}
	\author{ Yuri Kabanov \and Aleksei Kozhevnikov} 
	
	\institute{\at	Lomonosov Moscow State University, 
	Moscow, Russia, and  Universit\'e de Franche-Comt\'e, Laboratoire de Math\'ematiques, UMR CNRS 6623, 
	16 Route de Gray, 25030 Besan\c{c}on,  France \\
  \email{ykabanov@univ-fcomte.fr}. \\
	\and Lomonosov Moscow State University and ``Vega'' Institute, Moscow, Russia\\
	 \email{alexkozh27@gmail.com}}

\titlerunning{Optimal pair trading  with consumption}
\date{\today}
\maketitle

\begin{abstract}
We expose  a simple solution of the consumption-investment problem pair trading. 
The proof is based on the remark that the HJB equation   can be reduced to a  linear parabolic equation solvable explicitly.   
\end{abstract}

\smallskip

 

\section{Introduction}
This  note contains a short proof of the recent result  by Sahar Albosaily and Serguei  Pergamenchtchikov \cite{AP} on the consumption-investment optimal control problem in a pair trade setting. The pair trading is based on the idea that stocks of companies with the same business are strongly correlated and their difference fluctuates near zero.  
A  trader  matches a long position with a short position in two stocks having a high correlation. The portfolio value increment  is proportional to the increment of the spread between prices. By this reason such a setting, frequently  used  by hedge funds,  is also called spread trading.  
 The  mentioned paper \cite{AP} contains an extension  of the model considered earlier by Elena Boguslavskaya and Mikhail Boguslavsky in  \cite{BB} where the spread was modeled by the Ornstein--Uhlenbeck process and investor's  goal is to maximize only the expected utility of the  terminal wealth. The HJB equation in \cite{BB}, though looking rather involved, admits a solution which can be referred to as an explicit one. 
In \cite{AP}  the functional to optimize includes also  the  expected utility of consumption with the same power utility function.  The suggested analysis of the latter, rather involved and lengthy,  is based on a fixed point method. It leads to the optimal solution which is   the main contribution , see Ths. 5.1 and  5.3 in  \cite{AP}. Though it may happen that the ideas of \cite{AP}  could be useful in a more general context, they are not needed in the considered case. Here we provide arguments showing that the spread trading problem is not much more complicated than the classical Merton problem. The key ingredient of our proof is a reduction of the HJB equation  to a linear parabolic equation  admitting  explicit solution.

\section{Model}
First we recall briefly the formulation of  optimal control problem for spread trading.  Its dynamic on $[t,T]$ is 
given by the two-dimensional  process $(X_v,S_v)_{v\in [t,T]}$ with 
\bean
dX_v &=& (rX_v - \kappa_1 a_v S_v - c_v)dv + \alpha_v \sigma dW_v, \qquad  X_t=x,\\
dS_v &= &-\kappa S_vdv + \sigma dW_v,   \qquad \qquad \qquad \qquad \qquad\,  S_t=s.
\eean

The constants $r\ge 0$, $\sigma,\kappa>0$ are, respectively, the interest rate, market volatility, and mean-reverting parameter, and $\kappa_1:=r+\kappa$. The admissible control processes ${\bf u}:=(a,c)=(a_v,c_v)_{v\in [t,T]}$
 are predictable with respect to the filtration ${\bf F}^t$ formed by the $\sigma$-algebras $\mathcal{F}^t_v: = \sigma \{ W_u - W_t, \ t \le \theta \le v\}$ and   have trajectories in $L^2([t,T])\times  L^1_+([t,T])$. Moreover, the control process vanishes  after the instant when the component $X$ attains zero. The set of such controls is denoted by ${\cal A}_t$. Note also that $S$ does not depend on control, while $X=X^{\bf u}$ does.

The Bellman function of the problem is 
\begin{equation}
\label{ValueFunction}
J^*(x, s, t) := \sup_{\mathbf{u} \in \mathcal{A}_t}\mathbf{E} \Bigg[ \int_t^T c_v^\gamma dv + \beta X_T^\gamma  \Bigg] 
\end{equation}
where $0<\gamma<1$, $ \beta>0$. It is easily seen that the function $x\mapsto J^*(x, s, t)$ is concave and even homogeneous of order $\gamma$. 

We assume, as in \cite{AP}, that $\kappa\ge r$.

\section{Verification lemma}
The verification method  is the simplest way to find the solution of optimal control problem. It prescribes to consider the Hamilton--Jacobi--Bellman (HJB) equation 
\beq
\label{HJB0}
\sup_{(a,c)\in \bbr\times \bbr_+}H(x,s,t,a,c,z(x,s,t))=0, \quad z(x,s,T)=\beta x^{\gamma}, 
\eeq
where the  operator $H$ is given by the formula 
$$
H(x,s,a,c,z):=z_t+(rx - \kappa_1 a  - c)z_x-\kappa s z_s
 +\frac 12 \sigma^2 z_{ss} +a\sigma^2  z_{xs} + \frac 12 a^2\sigma^2 z_{xx} +c^\gamma.
$$

Let $z=z(x,s,t)\ge 0$ be a classical supersolution of the  HJB equation, that is  a function such that $H(x,s,a,c,z) \le 0$ for all $(a,c)$. Then $z$ dominates the Bellman finction $J^*$.  Indeed, take arbitrary  ${\bf u}\in \cA_t$.  By the Ito formula  for $\theta \in [t,T]$
$$
z(X^{\bf u}_\theta,S_\theta,t) + \int_t^\theta c^\gamma_vdv -  z(x,s,t)= 
M^{\bf u}_\theta +\int_t^\theta H(X^{\bf u}_v,S_v,a_v,c_v,z(X^{\bf u}_v,S_v,t))dv,
$$
where the stochastic integral 
$$
M^{\bf u}_\theta:=\int_t^\theta \big((rX^{\bf u}_v - \kappa_1 a_v  - c_v)z_x(X^{\bf u}_v,S_v,t)-\kappa S_v z_s(X^{\bf u}_v,S_v,t)\big) dW_v 
$$
as a process on the interval $[t,T]$ is a local martingale.  It is easily seen that $M^{\bf u}$ is bounded from below, hence, it is a  supermaringale and  $\E M^{\bf u}_\theta\le 0$ for every $\theta\in [t,T]$. It follows that 
$$
 z(x,s,t)\ge  \E \left[\int_t^T c^\gamma_vdv+\beta (X^{\bf u}_T)^\gamma\right]. 
$$
Since $u$ is arbitrary, this implies that  $z(x,s,t)\ge J^* (x,s,t)$. 
 
 The following assertion  is usually referred to as the verification lemma (or theorem). We use the following version.
 \begin{lemm} Let $z(x,s,t)\ge 0$ be a  solution of (\ref{HJB0}) which is $C^2$ in 
 $(x,s)$, $C^1$ in $t$, and concave in $x$. Let  ${\bf u}\in \cA_t$ be such that the process $H(X^{\bf u},S,a,c,z(X,S,t))\equiv 0$ (a.s.). Define the   family of random variables  $\cZ:=\{z(X_\tau,S_\tau,t)\}$ where  $\tau$ runs the set of  stopping times with values in $[t,T]$. If $\cZ$ is uniformly integrable, then ${\bf u}$ is the optimal control 
and  $z(x,s,t)= J^* (x,s,t)$.
 \end{lemm}

\section{HJB equation}
Note  that the supremum in $a$ and $c$ in the formula (\ref{HJB0})  is attained  at  
\beq
\label{sup}
\tilde a=\frac {\kappa_1 s z_x-\sigma^2 z_{xs}}{\sigma^2z_{xx}},   \qquad \tilde c=\Big(\frac{z_x}{\gamma}\Big)^{\frac 1{\gamma-1}}. 
\eeq
 Thus, the Cauchy problem (\ref{HJB0})  can be reduced to  the Cauchy problem  
\beq
\label{HJB}
z_t + \frac{\sigma^2}{2}z_{ss}- \frac{(\sigma^2 z_{xs} - \kappa_1 s z_x)^2}{2\sigma^2 z_{xx}}  + rxz_x - \kappa s z_s + (1-\gamma)\Big(\frac{z_x}{\gamma}\Big)^{\frac{\gamma}{\gamma - 1}} = 0
\eeq
with the terminal value $z(x,s, T) = \beta x^\gamma$. 

The equation above is  nonlinear but   a change of variable  tansforms it to a linear parabolic equation. Namely, we have 
\begin{lemm}
The solution of the terminal Cauchy problem for (\ref{HJB}) admits the representation $z(x, s, t) = x^\gamma u^{1-\gamma}(s, t)$ where $u(s,t)$ is the solution of the problem  
\beq
\label{EquationOnU}
u_t+\cL u+1=0, \qquad u(s,T)= \beta^{ \frac{1}{1-\gamma}},
\eeq
where 
\beq
\label{operator}
\cL u:=\frac{\sigma^2}{2}u_{ss}-\big(\gamma\kappa_\gamma+\kappa\big) su_s +\Big(\frac {1}{2\sigma^2}  \gamma \kappa_\gamma^2 s^2 + r \frac{\gamma}{1- \gamma} \Big)u
\eeq
with $\kappa_\gamma:=\kappa_1/(1-\gamma)$.
\end{lemm} 
{\sl Proof.}  
The substitution $z(x, s, t) = x^\gamma y(s, t)$ in (\ref{HJB})  leads to the problem
\beq
y_t +\frac{\sigma^2}{2}y_{ss}+\frac {\gamma}{1-\gamma} \frac{(\sigma^2 y_{s}/y - \kappa_1 s )^2}{2\sigma^2 }y    +\gamma r y - \kappa s y_s+(1-\gamma)y^{\frac{\gamma}{\gamma - 1}} = 0 
\eeq
with the terminal condition $y(s,T)=\beta$.

Let $y=u^{1-\gamma}$. Substituting the formulae  $y_t=(1-\gamma) u^{-\gamma}u_t$, $y_s=(1-\gamma) u^{-\gamma}u_s$,
$y_{ss}=(1-\gamma) u^{-\gamma}u_{ss}-\gamma (1-\gamma)u^{-1-\gamma}u_s^2$,  
$y_s/y=(1-\gamma)u_s/u$  in the above equation and dividing   both sides by $(1-\gamma) u^{-\gamma}$   we obtain the result. \fdem  

\medskip
The representation of $z$ given by the above lemma allows us to represent the formulae 
(\ref{sup}) in terms of the function $u=u(s,t)$ as follows:
\beq
\label{sup1}
\tilde a=x R(s,t) \  \hbox{where}\ R(s,t):=   \frac{u_s(s, t)}{u(s, t)}-s \frac{\kappa_\gamma}{\sigma^2},   \qquad \tilde c=\frac x{u(s,t)}. 
\eeq

\subsection{Explicit solution}
\begin{lemm} Let $g$ be a function satisfying the Riccati equation  
\beq
\label{eqg}
\dot{g} + \sigma^2g^2 - 2\big(\gamma\kappa_\gamma+\kappa\big) g+ \sigma^{-2}  \gamma \kappa_\gamma^2 = 0
\eeq
and let  $f$ be a function satisfying  the linear homogeneous equation
\beq
\label{eqf}
\dot{f} + \frac{\sigma^2}{2} g f + r \frac{\gamma}{1- \gamma} f = 0. 
\eeq  
Then the function $\tilde u(s,t):=f(t)e^{s^2g(t)/2}$ satisfies the equation $\tilde u_t+\cL \tilde u=0$.
\end{lemm}
{\sl Proof.} We have the following expressions: $\tilde u=f e^{s^2g/2}$, $u_s = sf g e^{s^2g/2}$, 
$$
\tilde u_t = \big(\dot{f} + (1/2)s^2\dot{g}f \big)e^{s^2g/2}, \quad \tilde u_{ss} =  \big(gf + s^2 g^2f \big)e^{s^2g/2}.
$$
Substituting them  into the formula (\ref{operator}) we get the result.  \fdem

\smallskip
Let $g=g^\theta$ and $f=f^\theta$ be two functions satisfying on $[0,\theta]$ the equations
(\ref{eqg}) and (\ref{eqf}) with the terminal conditions $g^\theta(\theta) = 0$ and 
$f^\theta(\theta) = \beta^{ \frac{1}{1-\gamma}}$. As an obvious corollary of the above lemma 
we get that the function $\tilde u^\theta (s,t):=f^\theta (t)e^{s^2g^\theta (t)/2}$ solves on $\bbr\times [0,\theta]$ the terminal Cauchy problem
\beq
\label{EquationOnUH}
\tilde u^\theta _t+\cL \tilde u^\theta =0, \qquad \tilde u^\theta (s,\theta )= \beta^{ \frac{1}{1-\gamma}}. 
\eeq

To alleviate formulae we skip $\theta$  when $\theta=T$. 
\smallskip

\begin{lemm} The function $\tilde u(s,t):=f(t)e^{s^2g(t)/2}$ solves the problem
\beq
\label{EquationOnUH}
\tilde u_t+\cL \tilde u=0, \qquad \tilde u(s,T)= \beta^{ \frac{1}{1-\gamma}}. 
\eeq
\end{lemm}

\begin{lemm}
Let $
h^\theta(s, t) := f^\theta(t) e^{s^2g^\theta(t)/2}$.
Then the function 
\beq
\label{Solution}
u(s, t): = \beta^{ \frac{1}{\gamma-1}} \int_t^T h^\theta(s, t) d\theta + \tilde u(s,t)
\eeq
solves (\ref{EquationOnU}).
\end{lemm}
{\sl Proof.} 
Note that 
$$
u_t(s,t) =  \beta^{ \frac{1}{\gamma-1}} \int_t^T h_t^\theta(s, t) d\theta -  \beta^{ \frac{1}{\gamma-1}} h^t(s, t) + \tilde u_t(s,t).
$$
and $h^t(s, t) =  \beta^{ \frac{1}{1-\gamma}}$. Since 
$$
u_t(s,t) +\cL u(s, t)+1= \beta^{ \frac{1}{\gamma-1}} \int_t^T(h_t^\theta(s, t)+\cL  h^\theta(s, t) )d\theta + \tilde u_t(s,t) +\cL \tilde u_t(s,t)=0,
$$
the result follows from the previous lemma.

\subsection{The Riccati  equation with constant coefficients}
Let $q=\sigma^2 g$. Then  $q$ solves the equation
\beq
\label{q}
\dot q+q^2 - 2(\gamma\kappa_\gamma+\kappa) q+ \gamma \kappa_\gamma^2 = 0, 
\eeq
Suppose that $\kappa>r\sqrt \gamma$. Then  the  quadratic equation $
 \lambda^2 - 2(\gamma\kappa_\gamma+\kappa)\lambda + \gamma \kappa_\gamma^2 = 0$
with the discriminant 
$$
D:=(\gamma\kappa_\gamma+\kappa)^2- \gamma \kappa_\gamma^2=
\left (\gamma \frac {\kappa+r}{1-\gamma}+\kappa\right)^2-\gamma \left(\frac {\kappa+r}{1-\gamma}\right)^2=\frac {\kappa^2-r^2\gamma}{1-\gamma}>0
$$
has the real roots  $\lambda_1:=\gamma\kappa_\gamma+\kappa+\sqrt D$ and $\lambda_2:=\gamma\kappa_\gamma+\kappa-\sqrt D>0$.  
Substituting   $q=p+\lambda_2$ into (\ref{q}) we obtain that    
$
\dot p+p^2 - Ap=0$
where $A:=2\sqrt D=\lambda_1-\lambda_2$. 
If  $p$ does not take the zero value, 
then $d(1/p)=-(1/p^2)dp$. The function  $P:=1/p$ satisfies the  equation  $\dot P=-A P+1$ and  can be represented as 
$$
P(t)=e^{-A(t-\theta)}\Big(P(\theta)+(1/A)(e^{A(t-\theta)}-1)\big)=(P(\theta) - (1/A))e^{-Atv-\theta)}+1/A. 
$$ 
Since $q=\lambda_2+1/P$, we obtain from here an explicit formula for $q$. 
In particular, if  $q(\theta)=0$, then  $P(\theta)=-1/\lambda _2$ and  for $t\in [0,\theta]$ 
$$
q^\theta(t)=\lambda_2-\lambda_2\frac{\lambda_1-\lambda_2}{\lambda_1 e^{(\lambda_1-\lambda_2)(\theta-t)}-\lambda_2}\ge 0, \qquad  q^\theta(\theta)=0. 
$$
It is easily seen that  
\beq
\label{boundq}
 \max_{0 \le t \le \theta \le T} q^\theta(t) = q^T(0) \le \lambda_2= \gamma\kappa_\gamma+\kappa -\sqrt{D}.
\eeq

In the case where $\kappa\ge r$  the discriminant $\sqrt D\ge \kappa$ and, therefore, 
\beq
\label{gT0}
g^T(0)=q^T(0)/\sigma^2\le \gamma\kappa_\gamma/\sigma^2.
\eeq

\subsection{Useful bounds}
In the sequel $C$ will denote a constant which value is no importance; it may be different  even in a chain of formulae. To simplify formulae we skip the dependence on $t$ where it has no importance. 

According to (\ref{eqf})  the function $f^\theta$ admits  an explicit expression    and we have that
$$
\beta^{ \frac{1}{1-\gamma}}\le f^\theta(t)=\beta^{ \frac{1}{1-\gamma}} \exp\left\{\int_t^\theta \Big(\frac{\sigma^2}{2}g^\theta(\nu)+\frac {r\gamma}{1-\gamma} \Big)d\nu\right\}\le C.
$$
It follows that 
\beq
\label{bounds-u}
\beta^{ \frac{1}{1-\gamma}}\le u(s,t)\le  Ce^{s^2 g^T(0)/2}\le Ce^{s^2 \gamma\kappa_\gamma/(2\sigma^2)}.
\eeq

\subsection{Uniform integrability}
Let us consider the process $X$ following on $[t,T]$ the stochastic differential equation  whose coefficients are defined in (\ref{sup1}):
$$
dX_v = X_v\big(r - \kappa_1 S_v R(S_v)  - 1/u(S_v) \big) dv + X_v\sigma R(S_v) dW_v, \qquad X^*_t=x.
$$
It can be given in more explicit way as $X_v=x\exp\{I_v+J_v\}$ where 
\beq
I_v:=\sigma \int_t^v R(S_\nu)dW_\nu, \qquad
J_v:=\int_t^v F(S_\nu)  d\nu.
\eeq
$$
F(S):=r - \kappa_1 S R(S)-(1/2) \sigma^2 R^2(S )- 1/u(S ).
$$

Our aim is to find $\delta>1$ such that $\sup_\tau \E z^\delta (X_\tau,S_\tau)<\infty$ 
where $\tau$ runs the set of all stopping times with values in $[t,T]$. 
Using the upper bound from (\ref{bounds-u}) we get that 
$$
z^\delta (X_\tau,S_\tau)\le CX^{\gamma\delta}_{\tau} u^{(1-\gamma)\delta}(S_{\tau})\le C
X^{\gamma\delta}_{\tau} e^{S_\tau^2 \delta\gamma\kappa_1/(2\sigma^2)}
$$
By the Ito formula applied to the square  Ornstein--Uhlenbeck process 
$$
S^2_\tau = s^2 + \int_t^\tau ( -2\kappa S^2_\nu + \sigma^2 ) d\nu + 2\sigma \int_t^\tau S_{\nu} dW_\nu, \quad v\in [t,T].
$$
Let $p>1$ and let $p'$ be its conjugate, i.e. $p':=p/(p-1)$. We prepare these numbers to use 
the H\"older inequality to isolate a stochastic exponential of a local martingale.  
With such a provision we rewrite the right-hand side of the above inequality as 
\beq
\label{prod}
z^\delta (X_\tau,S_\tau)\le C \exp\left \{\int_t^\tau G_\nu dW_\nu-\frac 12 p\int_t^\tau  
 G_\nu^2d\nu\right\}\exp\left \{\int_t^\tau \tilde G_\nu d\nu\right\}
\eeq
where we include in $C$ results of integration of bounded terms, 
\bean
G&:=&\delta\gamma\sigma R+(\delta \gamma \kappa_1/\sigma) S,\\
\tilde G &:=&
-\gamma\delta \kappa_1 S R-(1/2) \delta \gamma\sigma^2 R^2-(1/\sigma^2)\kappa \delta \gamma\kappa_1 S^2\\
&&+(1/2)p(\delta\gamma )^2\sigma^2 R^2+p\delta^2\gamma^2\kappa_1 RS
+(1/2)p(\delta \gamma \kappa_1/\sigma)^2 S^2
\eean
with the abbreviation $R:=R(S)$. 
If we take $p\delta\gamma=1$, then the coefficients at $R^2$ and $RS$ vanish and  
$$
\tilde G=\frac{\delta\gamma \kappa_1}{2\sigma^2} (\kappa_1-2\kappa) S^2=\frac{\delta\gamma \kappa_1}{2\sigma^2} (r-\kappa) S^2\le 0.
$$
Thus, the  $L^p$-norm of the first exponential in the rhs of (\ref{prod}) and the $L^{p'}$-norm of the second one are less or equal to one, $\E\big [ z^{\delta p}(X_\tau,S_\tau)\big]\le C$ and we get the needed uniform integrability property.

\section{Conclusion}
The consumption-investment problem in the setting of pair trade admits an explicit solution. 
The arguments are based on the observation that the HJB equation in the  Ornstein--Uhlenbeck spread model  is reduced to a linear parabolic PDE admitting an explicit solution. This observation   drastically simplifies the arguments in \cite{AP}.

\smallskip
{\bf Acknowledgement.} This work was supported by the Russian Science Foundation associated grants  20-68-47030 and 20-61-47043.

\end{document}